# DDoS Attack Detection Method Based on Network Abnormal Behavior in big data environment

Jing Chen[1], Xiangyan Tang[1, *], Jieren Cheng[1, 2] and Fengkai Wang[3], Ruomeng Xu[1]

[1]College of Information Science & Technology, Hainan University, Haikou 570228, China

[2]Key Laboratory of Internet Information Retrieval of Hainan Proince, Hainan Universty, Haikou 570228, China

[3] Waite Phillips Hall 3470 Trousdale Parkway, Los Angeles, CA 90089

* Corresponding Authors: Xiangyan Tang, tangxy36@163.com

**Abstract.** Distributed denial of service (DDoS) attack becomes a rapidly growing problem with the fast development of the Internet. The existing DDoS attack detection methods have time-delay and low detection rate. This paper presents a DDoS attack detection method based on network abnormal behavior in a big data environment. Based on the characteristics of flood attack, the method filters the network flows to leave only the "many-to-one" network flows to reduce the interference from normal network flows and improve the detection accuracy. We define the network abnormal feature value (NAFV) to reflect the state changes of the old and new IP address of "many-to-one" network flows. Finally, the DDoS attack detection method based on NAFV real-time series is built to identify the abnormal network flow states caused by DDoS attacks. The experiments show that compared with similar methods, this method has higher detection rate, lower false alarm rate and missing rate.

## 1    Introduction

Distributed denial of service (DDoS) attacks aim to quickly drain the communication and computing power of network targets by injecting large amounts of malicious traffic into them [1-3]. In recent years, DDoS attacks show a trend of increasing attack traffic, which can reach tens of GB or even hundreds of GB of attack bandwidth per second. Traditional defense technologies and mechanisms have been difficult to cope with. According to the cisco vision network index, global IP traffic was 1.2ZB per year or 96 exabytes per month by 2016. By 2022, global IP will reach 3.3ZB per year or 278 exabytes per month [4]. Due to the explosive growth and high complexity of network traffic in the environment of big data, the detection of network abnormal behavior has brought severe challenges, and a single defense method is difficult to cope with large-scale flooding attacks. The explosive growth of network flow due to the large data environment and the characteristics of high complexity, abnormal behavior detection to the network has brought serious challenges, single defense method is difficult to deal with a massive flooding attack successful defense mechanism, must be able to effectively filter out malicious traffic, and as small as possible to reduce the impact on the legitimate user traffic, can continue to respond quickly to threats, and has a small delay costs.

In order to overcome the above shortcomings, this paper proposes a flooding attack detection method based on abnormal network behavior in the big data environment. This method defines the fusion feature values (*NAFV*) to the network to identify abnormal behavior, the integral autoregressive Moving Average model is built based on *NAFV* (Auto Regressive Integrated Moving Average ARIMA) forecast classification model for real-time analysis of network traffic to identify the flood attacks, at the same time based on sliding window mechanism to *NAFV* sequence optimization computing resources, design the activation method based on threshold to reduce the computational overhead of attack detection. Experimental results show that compared with similar methods, this method has higher detection rate, lower false alarm rate and failure rate, and consumes less resources.

The rest of this article is arranged as follows. Section 2 introduces the relevant work. Section 3 introduces the determination of network abnormal behavior in detail. Section 4 is the experimental part. Section 5 is the conclusion.

## 2 Related Work

Software-defined networking (SDN) based DDoS attack detection technology detects anomalies through steps such as collecting network information, extracting analysis features and classification detection [5-7]. Yan et al. [8] discussed the new trends and characteristics of DDoS attacks in cloud computing, and comprehensively summarized the defense mechanism of DDoS attacks based on SDN, which is helpful to understand how to make full use of the advantages of SDN to overcome DDoS attacks in the cloud computing environment. Bawany et al. [9] studied and discussed in depth the detection and mitigation mechanism of DDoS attacks based on SDN, and classified them according to the detection technology. Taking advantage of the characteristics of SDN in network security, an active DDoS defense framework based on SDN is proposed. Yang et al. [10] proposed a scheduling based SDN controller architecture to effectively limit attacks and protect networks in DoS attacks. Kang et al. [11] proposed and proposed an active DDoS defense framework based on SDN based on the characteristics of SDN network security.

More and more distributed denial-of-service attacks are migrating to the cloud, posing serious challenges to the controllability of cloud computing and the security of the entire network space [12-14]. Li et al. [15] proposed and designed a DDoS attack source control method, PTrace (powerful trace), from the perspective of cloud service providers and in combination with the characteristics of cloud computing centers. PTrace controlled such attack sources from two aspects, packet filtering and malware tracing, to prevent the cloud from becoming a tool for DDoS attacks. Dick [16] used a combing method of security services called filter trees. The filter tree has five filters for detecting and resolving XML and HTTP DDoS attacks. Gao et al. [17] proposed a new classifier system for detecting and preventing DDoS TCP flooding attacks in public clouds（CS DDoS）. CS DDoS system provides a solution to protect stored records by classifying incoming packets and making decisions based on the classification results. Borisenko et al. [18] proposed a self-learning method that allows the detection model to adapt to network changes, which can minimize the error detection rate and reduce the possibility of marking legitimate users as malicious.

Existing models focus on DDoS attacks and victim attributes, but do not focus on botnet attributes, and botnet becomes the main technology of DDoS organization and management [19-21]. The key goal of distributed denial of service is to compile multiple systems and form botnets using infected zombies/agents over the Internet. The purpose is to attack a specific target or network with different types of packets. The infected system is controlled remotely by an attacker or a self-installed Trojan. Saied et al. [22] used artificial neural network algorithm to detect TCP, UDP and ICMP DDoS attacks, distinguished real traffic from DDoS attacks, and conducted in-depth training on the algorithm by using real cases generated by existing popular DDoS tools and DDoS attack modes. Ramanauskait et al. [23] proposed a DDoS attack model. The modeling results of different botnet allocation strategies show that the success of DDoS depends on attack dynamics. The proposed DDoS attack model can simulate different victim, attack and botnet characteristics, and apply these results to the game planning to analyze the probability of the victim's resistance to DDoS attack.

## 3 DDoS Attack Detection Method Based on Network Abnormal Behavior

### 3.1 Data Preprocessing

In order to eliminate the one-to-one and one-to-many network flow interference caused by normal flow and improve the detection rate of DDoS, the address correlation (The address correlation degree, ACD), IP flow feature (The IP flow features value, FFV is selected in this paper. IP stream interactive behavior feature (The IP flow interaction behavior feature, IBF), network flow multi-feature fusion algorithm (The IP flow multi-feature fusion, MFF), IP stream address semi-interactive anomaly degree (The IP flow address half interaction anomaly degree, HIAD), five feature extraction methods based on the above characteristics [24-28] are used to form the features of this paper. The reasons and the validity of features have been fully discussed in the author's original paper.

Definition 1. Assume that the network flow U is $< (t_1, s_1, d_1, p_1), (t_2, s_2, d_2, p_2), ..., (t_n, s_n, d_n, p_n) >$, $t_i$, $s_i$, in unit time t Di, pi respectively represent the time, source IP address, destination address, and port number of the i(i=1, 2, ..., n) packets. Classify the n data packets, that is, the data packets with the same source IP address and the destination IP address are grouped into the same class, and the class whose source IP address is Pi and the destination IP address is $P_j$ is $IP_{sd}$ ($P_i$, $P_j$).

For the class formed above, perform the following delete rule:

If there are different destination IP addresses $P_j$ and $P_k$ such that the classes $IP_{sd}$ ($P_i$, $P_j$) and $IP_{sd}$ ($P_i$, $P_k$) are not empty, delete all classes in which the source IP is $P_i$.

If the class formed by all the packets whose destination IP address is $P_j$ has only a unique class $IP_{sd}$ ($P_i$, $P_j$), delete the class of the packet whose destination IP address is $P_j$.

In order to process and compare a large number of IP addresses, this paper designs a database IPD that can quickly access IP addresses. This database can quickly mark, unmark and check whether a given IP is in the database in O (1) time.

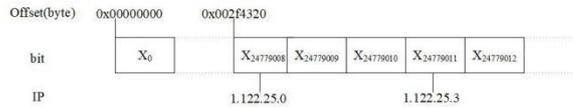

**Fig. 1.** IP address database example

For example, if there is an IPD data set as shown in Figure 3, if you want to check if the IP address 73.111.114.105 is marked, you can simply calculate the corresponding offset of the byte and bit. In the original data, the IP address 73.111.114.105 represents an unsigned integer 1232040553. In order to get the byte offset, divide 123240553 by 8 and the result is 154005069 in decimal, which is 0x092DEE4D in hexadecimal. The offset of the bit is the method of modulo, and the result of the calculation here is 1. Check the second bit of 0x092DEE4D. If set to 1, this IP address is marked as the old IP address, otherwise it is the new IP address.

$$Decimal\ IP\ address\ |8, byte \qquad (1)$$
$$Decimal\ IP\ address\ mod\ 8, bit \qquad (2)$$

IPD is not only applicable to IPv4, but can also map 128-bit IPv6 addresses to 32-bit integers using appropriate hashing algorithms. Because even if 25% of the hash table is used as the storage limit, it can still store $2^{30}$ different addresses.

### 3.2 Feature extraction

In unit time t, all IP packets of the network stream taken out from the normal network stream U are denoted as $I_k$. $I_k$ is filtered according to the above rules. The filtered sample group is denoted as $F_k$.

$$G_{I_j} \in F_K \text{ if } S_{I_{k_j}} \text{ is Many-to-one network flow} \quad \forall I_{k_j} \in I_k \tag{3}$$

Mark the IP address in $F_k$ as $O'$, all the IP addresses in $O'$ are old users. When k = 1, the maximum number of users is $O'_{max} = max(O'_{max}, ||F_k \cap O'||)$. After each unit time t, $F_k$ is merged into $O'$ to obtain the maximum number of old IP Numbers in the current unit time t[29-30].

$$O'_{max} = \begin{cases} ||F_1||, & k=1 \\ max(O_{max}, ||F_k \cap O||), & k \geq 2 \end{cases} \quad \forall F_k \tag{4}$$

The number of new users $N_k = ||F_k|| - ||F_k \cap O'||$. The set $F_k \cap O'$ represents the old user, the set $F_k \setminus O'$ represents the new user in the time period, and $D_k$ is the number of new users in the time period.

After processing the normal network flow U, which is a training sample $O'_{max}$, the obtained parameter $O'_{max}$ indicates the maximum number of old users in a $\Delta t$ period. The average number of new users $\overline{N}$ per unit time t can be calculated.

$$\overline{N} = \frac{\sum_{i=1}^{k} N_i}{k} \tag{5}$$

Using the same unit time t to sample the detection stream V, in each sampling area, we use a dictionary $W_k$ to store the number of visits to each source IP address $S_k$ during that time period. $W_k = [S_{k,i}, o_{k,i}]$ In this paper, we make the IP address set for all the source IP addresses in the k time period, so that we can represent the number of visits of the source IP $S_{k,i}$ in the k time period. In each unit time t, we calculate four eigenvalues, which are defined as follows.

Definition 2. *N* represents the percentage of old users in the current unit time t that exceed the maximum value of an old user on a time slice.

$$N = ||L_k \cap O'|| - O'_{max} - 1 \tag{6}$$

Then, we used $\Delta t$, $O'$, $O'_{max}$ and $\overline{N}$ four parameters for quantitative analysis of the testing process. The detection stream V is sampled using the same unit time t, and the dictionary $M_k$ is used to save all the source IP addresses of the unit time, $M_k = [S_{k,i}, O_{k,i}]$. S represents the keyword, and the time of occurrence is the value. Let $M_k[S_{k,i}]$ represent the visit times of source IP $S_{k,i}$ in the k time period. Let $L_k$ represent all values of $M_k$. At the end of each sampling time and time t, the fusion eigenvalues proposed in this paper are calculated. The detailed explanations of the four eigenvalues are as follows.

Definition 3. *A* represents the percentage of change of new users relative to the average number of new users.

$$A = \frac{||L_k \setminus O'|| - \overline{N}}{\overline{N}} \tag{7}$$

Definition 4. *F* represents the ratio of the maximum value of the current new user to the old user.

$$F = \begin{cases} \frac{||L_k \setminus O'||}{O'_{max}+1}, & if \ ||G_k \setminus O|| \neq 0 \\ \frac{||L_k \setminus O'||-1}{O'_{max}+1}, & otherwise \end{cases} \tag{8}$$

Definition 5. *V* represents the current access rate of the new user.

$$V = \frac{\sum\{W_k[S_{k,j}] | \forall S_{k,j} \in (G_k \setminus O')\}}{||G_k \setminus O'|| \Delta t} \tag{9}$$

Definition 6. The product of $N$, $A$, $F$ and $V$ is calculated on the basis of their definitions.

$$NAFV = -N \times A \times F \times V \tag{10}$$

Filter out all IP addresses that do not conform to the flood attack characteristics to ensure accuracy. The network is classified according to fused eigenvalues:

Normal flow. In general, the current number of old users $||G_k \cap O'||$ is close to the maximum value of the old user $O'_{max}$, making $O'$ close to ±0. The current number of new users is close to the average value of the new user. As a result, there will be a value of $A$ close to 0. For $F$, there are more old users than new users, and $F$ is a number close to 0. The access rate $A_k$ per second for each user is a small fixed value c. Multiply the four by minus. The final result will be very close to ±0.

$$\because N \to \pm 0, A \to 0, F \to +0, V = c$$
$$\therefore NAFV = -N \times A \times F \times V \to \pm 0$$

For sites for local users, there may be periods of low access, where the number of older users will decrease accordingly. The $O'$ value is close to -0. 5 or worse. 1. But this scenario does not affect the final model, $A$ and $F$ are still close to 0.

If the DDoS attack is taking place in DDoS, the current number of new users $||G_k \setminus O'||$ should be far larger than the average value of the new user, and the $D$ value will increase. If the DDoS attack is effective, the website or network can hardly provide service to the old user. The current number of old users $||G_k \cap O'||$ should be a very small value, such that the value of $O'$ is close to -1. The access rate of the new user $V$ is a large number.

$$\because N \to -1, A \gg 1, F > 0, V \gg 1$$
$$\therefore NAFV = -N \times A \times F \times V \gg +1$$

In addition, $N$ can represent the impact of DDoS attacks on older users, while $NAFV$ will be used to measure the impact of overall DDoS attacks: formula (10) can be interpreted as the overall impact value of DDoS.

Network congestion. Another detectable network flow is network congestion. When a hot topic arises, the number of new users and the number of old users will increase significantly. There are three characteristics: first, because of a large number of new users, $N$ should be very positive; Secondly, because of the wide range of hot topics, the old users are likely to access it, so the $N$ value should be higher than 1. Finally, even if there are many new users, the constant c will be a smaller value because it follows the TCP/IP protocol with the normal users.

$$NAFV = -N \times A \times F \times V \ll -1$$

$F$ indicates the degree of interest of external users to a particular hot topic: if $R_k \in (0, 1]$ it is a particular hot topic, it can imply that older users are more concerned with the topic; otherwise, it implies that new users are more concerned about the topic.

**3.3 Classification Model Based on NAFV**

After filtering the data set, only "many-to-one" network traffic is left, which improves the accuracy of prediction. Based on the noiseless data set and fused eigenvalues, the trained ARIMA model is applied to the trend prediction. After $N$ times of unit time sampling and calculation of fused eigenvalues, the time series K=$NAFV$, $i$=1, 2, ... and $N$ are obtained. By observing the autocorrelation coefficient and partial autocorrelation coefficient as shown in figure 3 and 4, according to the red pool Information Criterion (Akaike Information Criterion) ARIMA model parameters, this paper models should be ARIMA (2, 2, 1). In this paper, ARIMA model is set up in R language, because this article the source data is smooth, this article uses the second-order difference processing raw $NAFV$ data, can see the forecast and actual data as shown in figure 4 high fitting degree, and smoothly passed the ADF test.

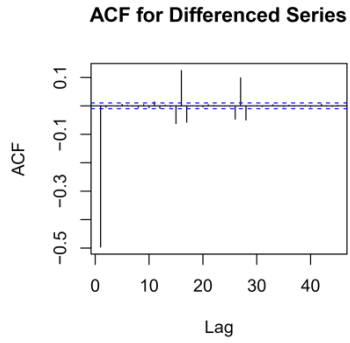 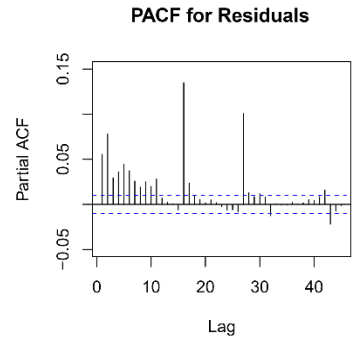

     **Fig. 2.**   ACF on time series K           **Fig. 3.**   Partial ACF on time series K

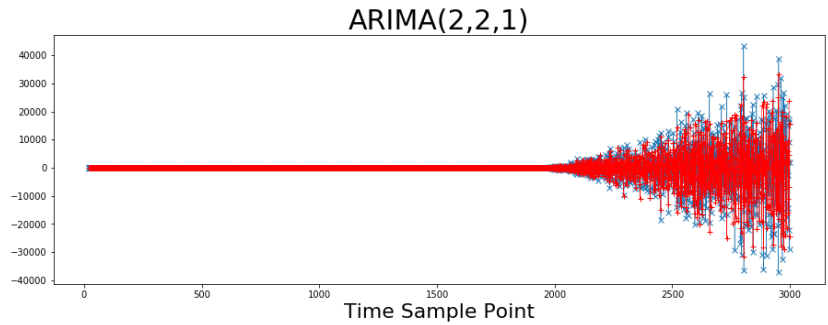

**Fig. 4.**   ARIMA (2,2,1)

In order to further verify the model in this paper, Ljung test is used to detect whether the standard residual is white noise. Time series, autocorrelation and partial autocorrelation of standard residuals are shown in figure 5 and 6, respectively.

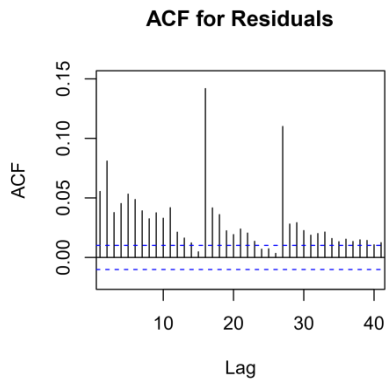 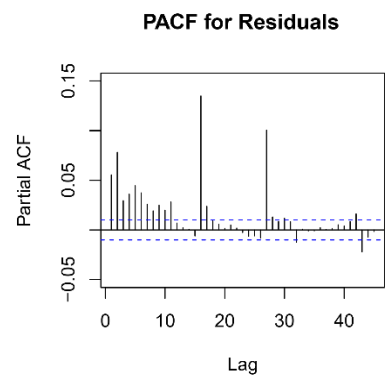

    **Fig. 5.**   ACF of standardized residuals         **Fig. 6.** ACF of standardized residuals

In this paper, a network flow detector based on threshold was used. When the abnormal value of *NAFV* was detected, the trained ARIMA model was used for trend prediction. When continuous samples exceeded the threshold value a, the alarm would be triggered. According to the experimental observation, in this paper, the threshold value $a=25$, $\beta=2$, that is, when there are 2 continuous abnormal sampling points. In the ARIMA trend prediction module, a sliding window mechanism is added, which contains the nearest w points. In the *NAFV* points predicted by the ARIMA model, the number of abnormal *NAFV* values is denoted as, and if $\frac{\gamma}{w}$ exceeds a certain percentage, a DDoS warning will be activated. Otherwise, the current network stream is continuously detected until either $\frac{\gamma}{w}$ is 0 or the criteria are met.

    In the big data environment, it is required not only to detect DDoS attacks quickly and accurately,

but also to consume computing resources as little as possible to avoid or minimize interference to the normal network flow. In this paper, we design a method to activate trend prediction. The ARIMA trend prediction module is inactive in normal flow, and it is triggered only after the detection of continuous abnormal points. Finally, when the ARIMA trend prediction module determines that there is a DDoS attack or that the stop condition is met, it is paused to save computing resources.

This paper uses normal network flows as training samples to generate an IP address database IPD, for recording the IP addresses of older users. The basic parameters of the algorithm and the parameters of the ARIMA model are generated at the same time, and the IPD is used to determine whether an IP address is an old user. After each unit of time t, the value of the fused feature *NAFV* is calculated and sent to the network flow recognizer as a sampling point. The recognizer has a preset threshold $\alpha$. When the *NAFV* value of a sampling point exceeds the threshold $\alpha$, the sampling point is marked as an outlier. When there are continuous $\beta$ outliers, The ARIMA module is started for trend prediction. The overall framework is shown in figure 7.

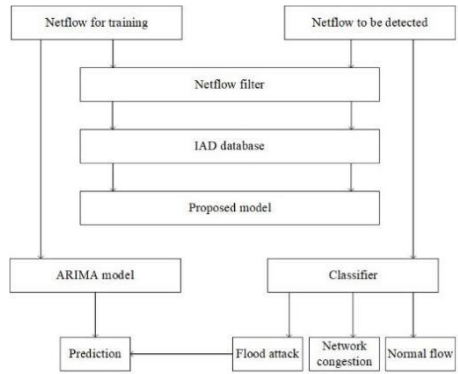

**Fig. 7.** Framework of the proposed method.

## 4. Experiment and Analysis of Results

### 4.1 Experimental Datasets and Evaluation

As described in the CAIDA DDoS attack 2007 dataset [30], the duration of each network stream is 5 minutes, the entire network stream starts at 13:49:36, and the DDoS attack occurs at approximately 14:15:56. So track the beginning of a DDoS attack in the detection stream. Select $\Delta t = 0.8$ as the testing flow sample time, and calculate the corresponding *NAFV* value, results shown in figure 8.

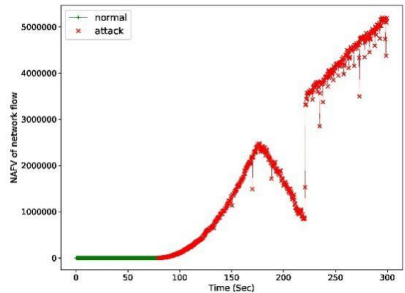

**Fig. 8.** Ground truth of CAIDA DDoS attack dataset 2007

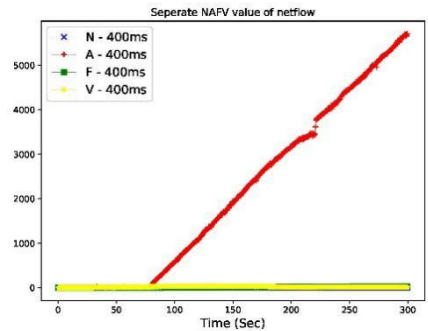

**Fig. 9.** Separate NAFV of network flow

After analyzing the feature values, it can be seen that the variable N changes the most, followed by other variables. In order to better detect attacks, the variable weight of fusion eigenvalues is assigned as follows:

$$NAFV = -\omega_1 N \times \omega_2 A \times \omega_3 F \times \omega_4 V \ , \ \omega_1 + \omega_2 + \omega_3 + \omega_4 = 1 \quad (11)$$

Where $w_1$, $w_2$, $w_3$ and $w_4$ represent the weight of the four features respectively by calculating from principal component analysis in Section 3.2

In order to evaluate the effectiveness of the proposed algorithm, TN is defined in this paper as the number of accurately identified DDoS network stream samples, FN is the number of normal stream but wrongly marked as DDoS network stream, TP is the number of correctly identified ordinary users, and TN is the number of DDoS network stream samples wrongly identified as normal stream samples.

Detection rate DR. Indicates the ability of the classifier to identify real DDoS attack streams.

$$DR = \frac{TN}{TN+FN}$$

False negatives rate MR. Indicates the possibility that the classifier cannot distinguish between true DDoS attack streams.

$$MR = \frac{FN}{TN+FN}$$

False positive rate FR. Explain the possibility that a normal user is marked as an attacker by a classifier error.

$$FR = \frac{FP}{TP+FP}$$

**4.2    Analysis based on fusion feature values**

Based on a good training data set and the proposed fusion eigenvalue *NAFV*, ARIMA model is further applied to the prediction work. It shows good results in both sample predictions and future predictions. The experiment in this paper is based on CAIDA DDoS attack 2007 data set, and different t = [0.04,0.08,0.4,0.8,1.6] is used as the parameter training model. Firstly, we use the original data flow to calculate the network feature values according to formula (10) for prediction. Secondly, we use the filtered data flow to calculate the network feature values according to formula (11) for prediction. Both tests can successfully distinguish the normal flow from the DDoS flow.

When the sampling time is small, false alarm will occur in the later period of attack. When the sampling time is large, false alarm will occur in the early period of attack, as shown in figure 10. The experimental data and the characteristics of the proposed algorithm are analyzed as follows. When t is small the initial stage of a DDoS attack may not have many new users, or a relatively large number of old users in a short period of time, even if the ARIMA model takes past data into account, such a situation will still lead to detection failure. When t is large, in the early stage of attack, the change quantity of new users and old users is not very large, resulting in false alarm.

In the case of setting weights for variables, the false alarm rate and false alarm rate are greatly reduced, and the accuracy of prediction is improved.

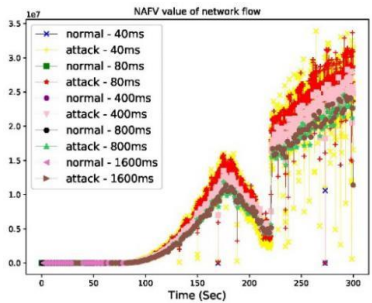    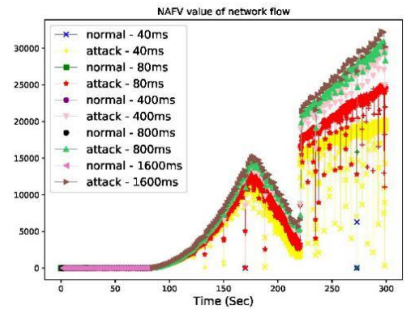

**Fig. 10.** NAFV value of network flow           **Fig. 11.** NAFV value of network flow

### 4.3 Comparative Analysis of Various Testing Methods

This paper compares the proposed method with existing methods, including SMPM method, c-svr regression prediction method and ARIMA method. With these methods, the Detection experimental results of unfiltered network traffic are Detection rate1, Missing rate1 and False alarm rate1, and the Detection experimental results of filtered network flow are Detection rate2, Missing rate2 and False alarm rate2. In order to evaluate the effectiveness of the proposed algorithm, TN is defined in this paper as the number of accurately identified DDoS network stream samples, FN is the number of normal stream but wrongly marked as DDoS network stream, TP is the number of correctly identified ordinary users, and TN is the number of DDoS network stream samples wrongly identified as normal stream samples.

Table. 1.   comparison of methods

| Method | Detection rate1 | Detection rate2 | Missing Rate1 | Missing Rate2 | False alarm Rate1 | False alarm Rate2 |
|---|---|---|---|---|---|---|
| Proposed method | 99.83% | 99.95% | 0.17% | 0.04% | 0.01% | 0% |
| c-SVR | 95% | 96.21% | 4.89% | 3.35% | 7.966% | 5.43% |
| SMPM | 99.6% | 99.71% | 0.4% | 0.28% | 8.89% | 8.72% |
| ARIMA | 90% | 91.18% | 9.99% | 9.04% | 5.973% | 4.22% |

1) Detection rate. Represents the ability of a classifier to recognize a real DDoS attack stream.

$$DR = \frac{TN}{TN+FN}$$

2) Missing Rate. Indicates the possibility that the classifier cannot distinguish the real DDoS attack stream.

$$MR = \frac{FN}{TN+FN}$$

3) False alarm Rate. Indicates the probability that a normal user is incorrectly labeled as an attacker by the classifier.

$$FR = \frac{FP}{TP+FP}$$

## 5   Conclusion

In this paper, the network traffic is filtered according to the characteristics of flood network flow, which reduces the overhead of experimental operation and improves the accuracy of attack detection. The fusion eigenvalue *NAFV* is defined to describe the change characteristics of network IP address, and the network flow is classified. The fusion eigenvalue *NAFV* is defined to describe the change characteristics of network IP address, and the network flow is classified. Based on *NAFV*, ARIMA predictive classification model is established to identify DDoS attacks. Experimental results show that the proposed algorithm can detect DDoS attacks more accurately and efficiently than similar methods. In the following work, we will further study how to add reinforcement learning methods to the existing framework to further improve the detection accuracy. With the repaid development of cloud robotics [32] and smart cities [33], the method will be widely used.

**Funding**

This work was supported by the Hainan Provincial Natural Science Foundation of China [2018CXTD333, 617048]; National Natural Science Foundation of China [61762033, 61702539]; Hainan University Doctor Start Fund Project [kyqd1328]; Hainan University Youth Fund Project [qnjj1444].

This work is partially supported by Social Development Project of Public Welfare Technology Application of Zhejiang Province [LGF18F020019]